\documentclass[aps,apl,twocolumn]{revtex4}
\usepackage{graphicx}
\usepackage{dcolumn}
\usepackage{bm}
\usepackage {epsfig}
\usepackage {graphics}

\usepackage {color}

\begin{document}

\title{Pressure effects on the superconducting thin film Ba$_{1-x}$K$_{x}$Fe$_{2}$As$_{2}$}

\author{Eunsung Park, Nam Hoon Lee, Won Nam Kang, and Tuson Park}
\email[]{tp8701@skku.edu}
\affiliation{Department of Physics, Sungkyunkwan University, Suwon 440-746, Korea}

%\date{\today}

\begin{abstract}
We report electrical resistivity measurements on a high-quality Ba$_{1-x}$K$_{x}$Fe$_{2}$As$_{2}$ thin film ($x=0.4$) under pressure. The superconducting transition temperature (=39.95~K) of the optimally-doped thin film shows a dome shape with pressure, reaching a maximal value 40.8~K at 11.8~kbar. The unusually high superconducting transition temperature and its anomalous pressure dependence are ascribed to a lattice mismatch between the LaAlO$_3$ substrate and the thin film. The local temperature exponent of the resistivity ($n=d\text{ln}\Delta\rho/d\text{ln}T$) shows a funnel shape around the optimal pressure, suggesting that fluctuations associated with the anomalous normal state are responsible for high-temperature superconductivity.
\end{abstract}

%\pacs{}

\maketitle

Since the report of superconductivity in F-doped LaFeAsO$_{1-x}$F$_{x}$~\cite{hosono08}, Fe-based superconductors have attracted strong interest because of their high superconducting transition temperature (high-$T_c$) and the size of the materials space where superconductivity is discovered~\cite{stewart11}. Similarity to the cuprate superconductors has been pointed out in that superconductivity is introduced when carriers are induced by hole/electron doping the antiferromagnetic parent compounds and FeAs or FeSe layers are deeply involved in the superconductivity as the CuO layers are in the high-$T_c$ cuprates. It has been recently suggested that the anion height measured from the Fe plane is an important parameter that controls the Fermi surface topology and consequentially the superconductivity~\cite{clee08, kuroki11}. There exists an optimal height ($h_m$) where the superconducting transition temperature is a maximum and $T_c$ decreases as the height departs from $h_m$.

Thin films grown on different types of single crystal substrates may manifest uniaxial pressure effects because a lattice mismatch relative to the substrate leads to either contraction or expansion of the film and a change in the anion height~\cite{hiramatsu12}. Iida $et$ $al.$ reported that $T_c$ of a Ba(Fe,Co)$_2$As$_2$ (Co-Ba122) thin film changes from 16.2 to 24.5~K with increasing $c/a$, where the out-of-plane lattice parameter $c$ decreases with an increase of the in-plane lattice constant $a$ for different types of substrates~\cite{Iida09}. Bellingeri $et$ $al.$ found that the $T_c$ of 21~K for a FeSe$_{0.5}$Te$_{0.5}$ thin film is significantly higher than 16.2~K for a bulk crystal, suggesting that strain effects in thin films may provide a way to raise $T_c$~\cite{bellingeri10}. 

Despite the ever growing interest in thin film studies, however, the high volatility of the doping elements often makes it difficult to fabricate thin films of the iron-pnictide superconductors. An electron-doped BaFe$_{2-x}$Co$_x$As$_2$ thin film with $T_c \approx 20$~K has been synthesized with relative ease because cobalt atoms have a lower vapor pressure, while a hole-doped Ba$_{1-x}$K$_x$Fe$_2$As$_2$ thin film has been fabricated only recently by overcoming the technical difficulty of a large difference in the vapor pressures of constituent elements~\cite{slee09, slee10, Iida10, katase10, takeda10, nlee10}. Lee $et$ $al.$ recently reported successful synthesis of a thin film of Ba$_{0.6}$K$_{0.4}$Fe$_2$As$_2$, where $T_c$ is as high as 40~K~\cite{nlee10}. Here we report for the first time the dependence of superconducting properties on externally applied pressure for a Ba$_{0.6}$K$_{0.4}$Fe$_2$As$_2$ thin film. Unlike single crystalline compounds, $T_c$ of the optimally doped thin film does not decrease monotonically with pressure but rather shows a dome centered around 11.8~kbar (=$P_c$). The electrical resistivity of the film in the normal state shows an anomalous power-law temperature dependence under pressure. The local temperature exponent of the resistivity shows a funnel shape above the optimal pressure $P_c$, suggesting that fluctuations relevant to anomalous normal state properties are responsible for high-$T_c$ superconductivity in the Fe-pnictides superconductors. 

Optimally doped thin films of Ba$_{0.6}$K$_{0.4}$Fe$_{2}$As$_{2}$ were grown on LaAlO$_{3}$ (001) substrates by a pulsed laser deposition (PLD) method, and X-ray diffraction measurements showed that films with a tetragonal structure grow preferentially along the crystalline c-axis~\cite{nlee10}. The lattice parameters determined from a least-squares refinement of the x-ray pattern are $a=3.9068$~\AA~and $c=13.4037$~\AA, where $c/a=3.4308$ is slightly larger than that of optimally doped bulk crystals (=3.40)~\cite{rotter08}. Electrical resistivity of the K-doped BaFe$_{2}$As$_{2}$ thin film was measured by a standard four-probe technique in a closed cycle refrigerator (CCR). A clamp-type Be/Cu hybrid cell with a NiCrAl alloy insert was used for resistivity measurements up to 28.2~kbar, and a change in the resistivity of an annealed manganin wire was used as a pressure monitor at room temperature~\cite{dmowski99}.

\begin{figure}[tb]
\includegraphics[width=7cm,clip]{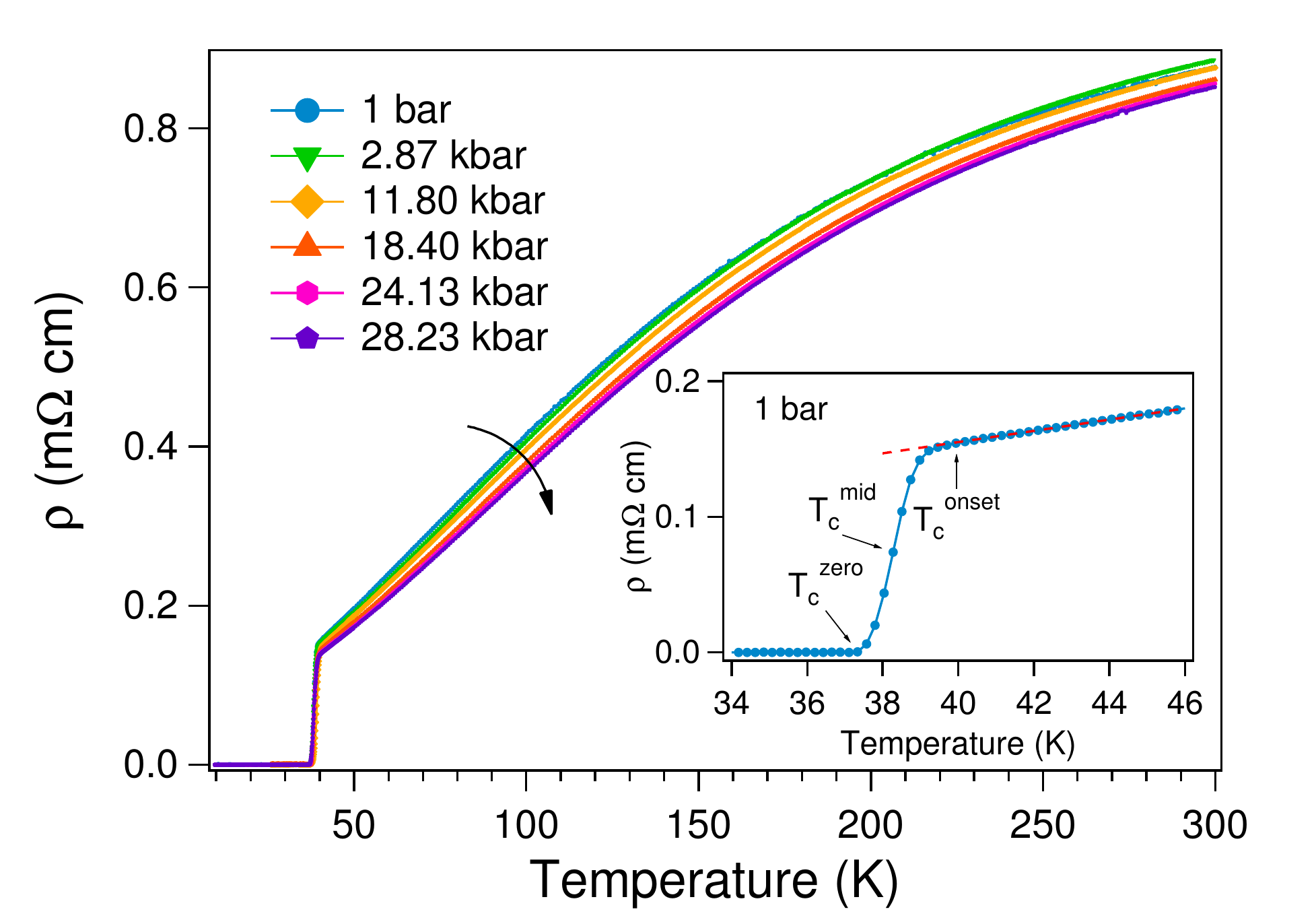}
\caption{\label{fig_1}(color online) Electrical resistivity of a Ba$_{0.6}$K$_{0.4}$Fe$_{2}$As$_{2}$ thin film deposited on a LaAlO$_{3}$~(LAO) substrate under pressure. The arrow indicates a pressure increase from 1~bar to 28.23~kbar. Inset: The resistivity at ambient pressure is displayed near SC transition temperature.}
\end{figure}
Figure~1 shows the temperature dependence of the in-plane electrical resistivity $\rho$ of a thin film of Ba$_{0.6}$K$_{0.4}$Fe$_2$As$_2$ under pressure. As shown in the inset, the onset temperature of a superconducting (SC) phase transition ($T_c^{onset}$), which is defined as the deviation point from a linear temperature dependence of the resistivity in the normal state (dotted line), is 39.95~K at ambient pressure, the highest for optimally doped films on a LaAlO$_3$ (LAO) substrate. The SC transition width, the difference between a 10~\% to 90~\% drop of the normal state resistivity at $T_c ^ {onset}$, is 1.30~K and the residual resistivity ratio (RRR) $\rho(300~\text K)/\rho(0~\text K)$ at 1~bar is 38, indicating the high quality of this thin film. Here, $\rho (0~\text K)=0.023~ m\Omega$cm is an extrapolated value from the $\rho = \rho(0~\text K) + AT^{1.2}$ dependence in the normal state. Externally applied pressure monotonically reduces the resistivity in the normal state because of an increased overlap among adjacent ligand orbitals under pressure: $\rho$ at 41~K and 300~K decreases at a rate of -0.6 and -1.1~$\times 10^{-3}~m\Omega$cm/kbar, respectively.

\begin{figure}[tb]
\includegraphics[width=7cm,clip]{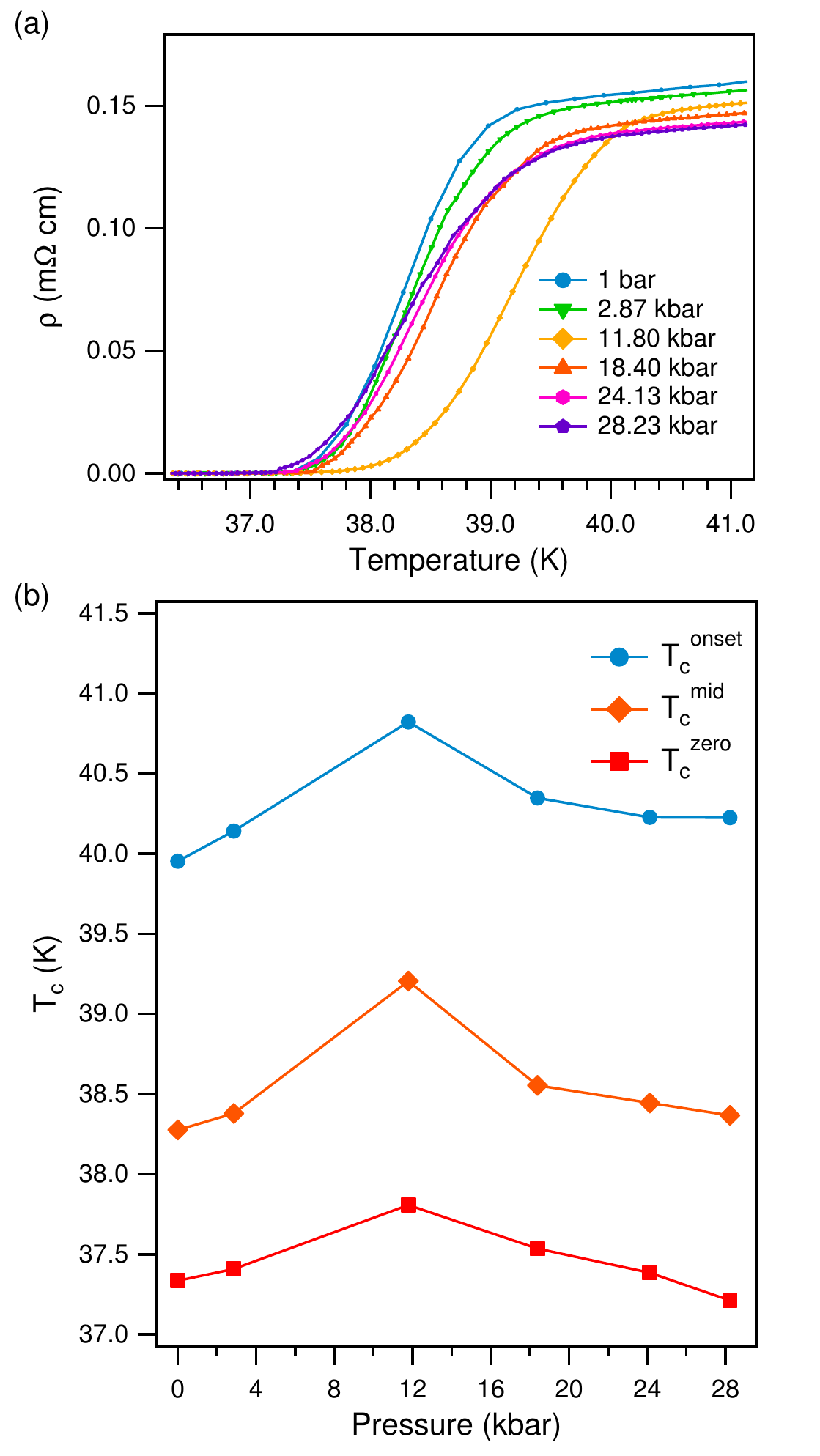}
\caption{\label{fig_3a,fig_3b}(color online). (a) Pressure dependence of the resistivity of a Ba$_{1-x}$K$_{x}$Fe$_{2}$As$_{2}$ $(x=0.4)$ thin film near its superconducting transition temperature. (b) Superconducting transition temperature vs pressure phase diagram, where $T_c^{onset}$, $T_c^{mid}$, and $T_c^{zero}$ represent the onset, midpoint, and $\rho =0$ SC transition temperatures, respectively.}
\end{figure}
The pressure evolution of superconducting transition temperatures is plotted in Fig.~2, where $T_c^{onset}$, $T_c^{mid}$, and $T_c^{zero}$ represents the onset, midpoint and zero-resistance transition temperatures, respectively (see the inset of Fig.~1). $T_c^{onset}$ gradually increases from 39.95~K at 1~bar to 40.8~K at 11.80~kbar and then decreases with further increasing pressure, showing a dome-shaped superconducting phase centered on the optimal pressure $P_c$(=11.80~kbar). Such pressure effects on the thin film are unexpected because $T_c$ of single crystals with similar stoichiometry is continuously suppressed with pressure~\cite{yamazaki10, torikachvili08}. For Ba$_{1-x}$K$_x$Fe$_2$As$_2$ bulk crystals, the $c/a$ ratio almost linearly increases with potassium concentration $x$ and is approximately 3.410 for an optimally K-doped compound ($x=0.4$)~\cite{rotter08}. The ratio $c/a=3.4308$ observed in our thin film corresponds to an overdoped concentration $x=0.46$ for a bulk crystal, ruling out the possibility that the thin film on LAO is underdoped. In addition, the $T_c$ of 39.95~K is too high for any underdoped compound. For comparison, $T_c$ of an optimally doped bulk crystal is 38~K~\cite{rotter08, luo08}.

The unusual pressure dependence of the optimally doped thin film may be ascribed to a high $c/a$ ratio that arises from a lattice mismatch between the film and LAO substrate. It has been reported that the $c/a$ ratio of a Ba$_{0.6}$K$_{0.4}$Fe$_2$As$_2$ thin film deposited on Al$_2$O$_3$ (AO)and LaAlO$_3$ (LAO) substrates is 3.4028 and 3.4308, respectively~\cite{nlee10}. The superconducting transition temperature of a film on a AO substrate is almost 1~K higher than that on a LAO substrate, indicating that the anion height of a AO-film is close to the optimal height $h_m$, while the anion height of a film on LAO is slightly larger than $h_m$. Lattice constants of the LAO substrate are almost independent of hydrostatic pressure and isotropic among crystalline axes: the relative change in lattice constants being only 0.55\% at 32.1~kbar~\cite{zhao04}. In contrast, the lattice constants of a tetragonal BaFe$_2$As$_2$ single crystal are compressed significantly both along the crystalline a- and c-axes with pressure: at 32.1~kbar, the lattice constant $a$ and $c$ decreases by 1.03 and 2.02\%, respectively~\cite{kimber09}. When the thin film of Ba$_{0.6}$K$_{0.4}$Fe$_2$As$_2$ is under hydrostatic pressure, it effectively experiences a uniaxial pressure along the c-axis because the change in the lattice constant $a$ is almost negligible because it is pinned to the substrate lattice, while the lattice constant along the c-axis is compressed as easily as in a single crystal. Consequently, hydrostatic pressure acts as an effective uniaxial pressure along the c-axis and controls the anion height of the film to the optimal $h_m$ at 11.8~kbar, therefore making the dome-shaped SC phase under pressure.

\begin{figure}[tb]
\begin{center}
\includegraphics[width=7cm,clip]{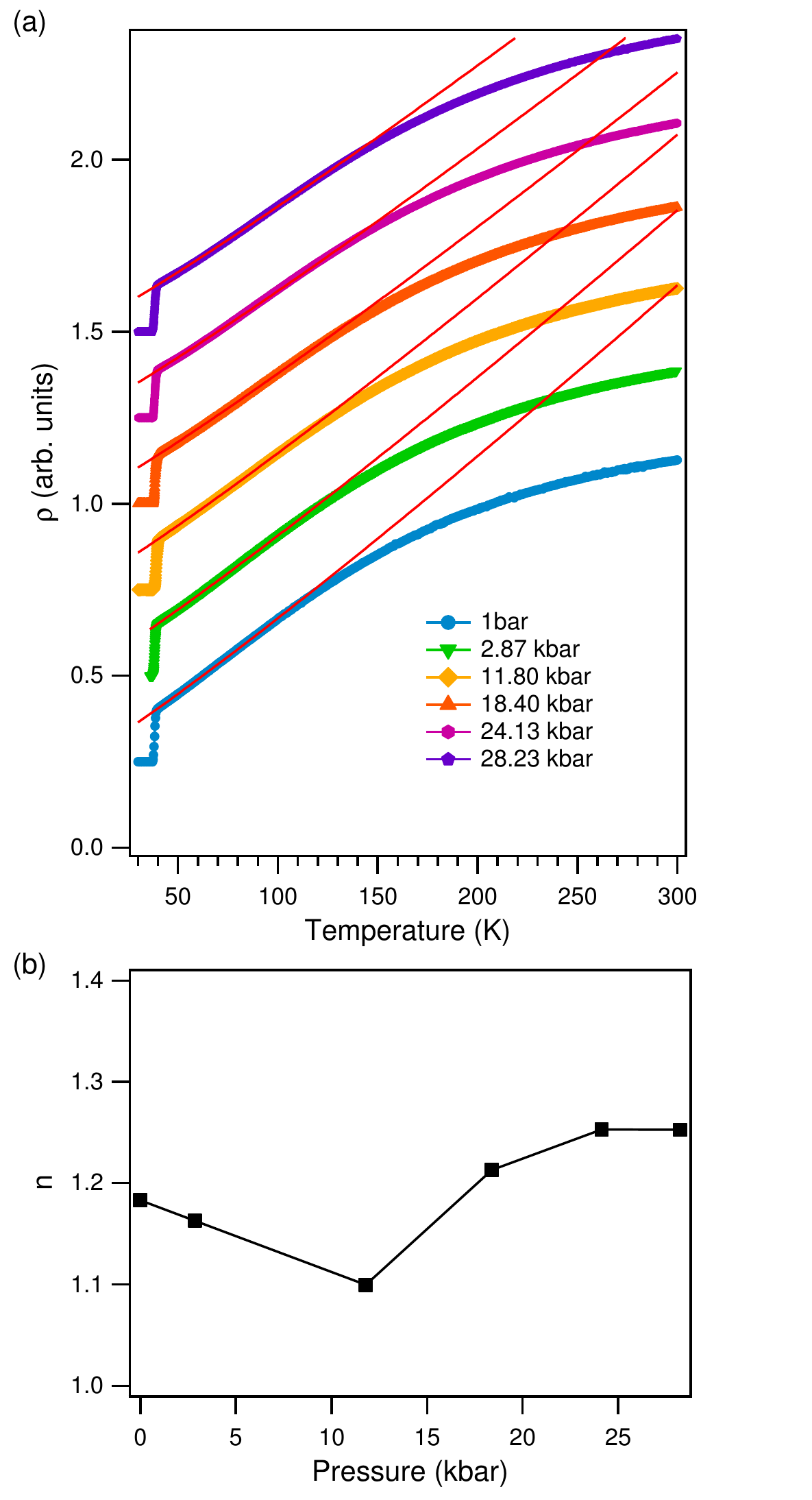}
\end{center}
\caption{(color online).(a) Temperature dependence of the resistivity of Ba$_{1-x}$K$_{x}$Fe$_{2}$As$_{2}$ $(x=0.4)$ under pressure and a least-squares fit of $\rho=\rho_{0}+AT^{n}$ (solid lines). The resistivity under pressure is shifted rigidly upwards for clarity. (b) Pressure dependence of the resistivity coefficient $n$ of the power-law form.}
\end{figure}
Figure~3(a) displays the temperature dependence of the resistivity of the Ba$_{0.6}$K$_{0.4}$Fe$_2$As$_2$ thin film on the LAO substrate for several pressures. Regardless of the applied pressure, a power-law form of $\rho=\rho_0+AT^n$ best describes the normal state resistivity over an extended temperature range above $T_c$, where the exponent $n$ of the resistivity changes between 1.1 and 1.25 under pressure (see Fig.~3(b)). The anomalous temperature exponent $n$ is consistent with previous work on single crystals that reported a gradual change of $n$ from 2~(undoped) to 1~(at optimal potassium doping)~\cite{shen11}.

Figure~4 plots the local temperature exponent of the resistivity ($n=d\text{ln}\Delta\rho/d\text{ln}T$) in $T-P$ phase space, where $\Delta \rho=\rho(T)-\rho(T=0K)$. The local exponent $n$ is less than 1.25 below 100~K in the measured pressure range, which is comparable to the results via a global fit of the resistivity (see Fig.~3(a)). The local analysis, however, reveals a delicate topology in the resistivity exponent, showing a funnel shape with $n=1.1$ above the optimal pressure. Even though the optimally doped Ba$_{0.6}$K$_{0.4}$Fe$_2$As$_2$ thin film at 1~bar is probably in the vicinity of a quantum critical point (QCP), external pressure sensitively tunes the film right to the QCP where the associated quantum fluctuations diverge. A similar analysis based on the local resistivity exponent has been performed to show evidence for quantum critical behavior in other iron-pnictides superconductors as a function of chemical doping~\cite{kasahara10}, which inherently incur disorder whose effects on the electron scattering may change with doping concentration. In this work, the K-doped Ba122 thin film with $x=0.4$ was sensitively tuned to a critical point via pressure, thus allowing us to study the quantum critical behavior without incurring additional disorder. A funnel shaped resistivity exponent above an optimal pressure has been also reported in the pressure-induced heavy fermion superconductor CeRhIn$_5$~\cite{park08}. Future characterization of the spectra of quantum fluctuations that are pertinent to the high SC transition temperature in iron-pnictides superconductors requires simultaneous tuning of such external parameters as pressure, magnetic field, and low temperatures.
\begin{figure}[tb]
\begin{center}
\includegraphics[width=7cm,clip]{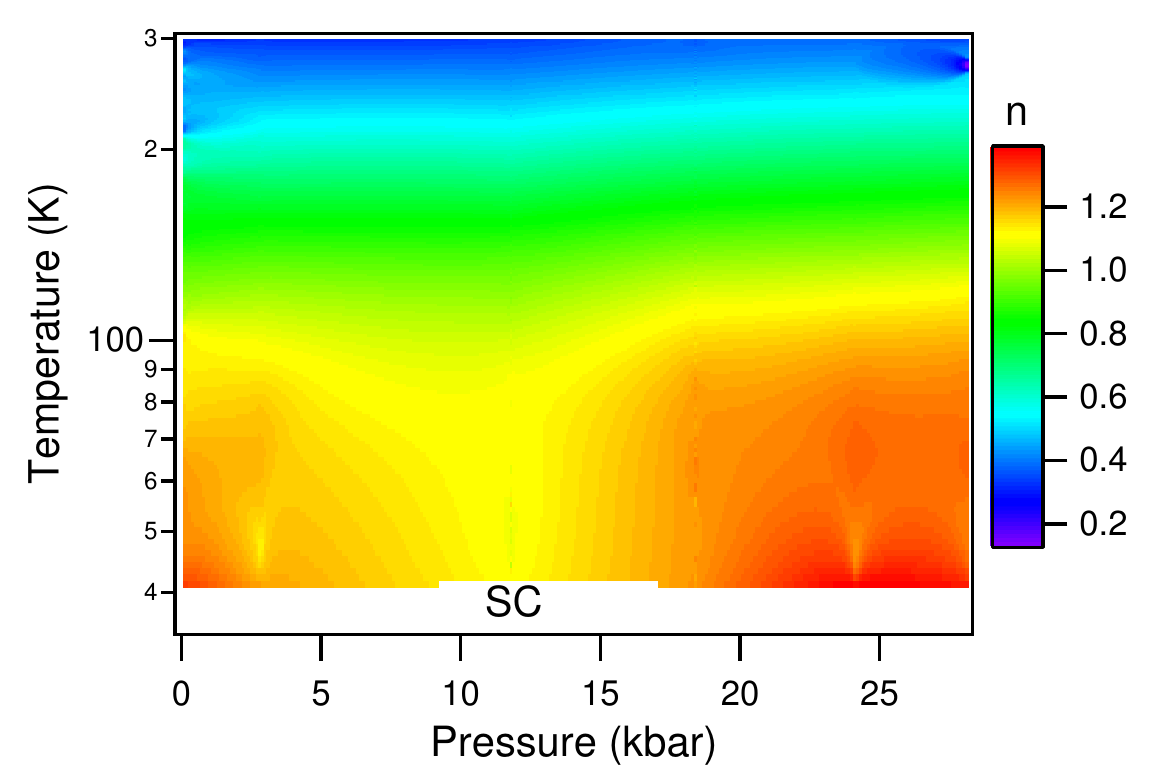}
\end{center}
\caption{\label{fig_4}(color online). Evolution of the local temperature exponent $n$ of a Ba$_{0.6}$K$_{0.4}$Fe$_{2}$As$_{2}$ thin film is plotted in the temperature-pressure phase space. The local exponent is defined as $n=d\text{ln}\Delta\rho/d\text{ln}T$, where $\Delta\rho=\rho(T)-\rho(T=0~\text K)$.}
\end{figure}

We fabricated Ba$_{1-x}$K$_{x}$Fe$_{2}$As$_{2}$ $(x=0.4)$ thin films on a LaAlO$_3$ substrate, where $T_c$ is 39.95~K. The sharp superconducting phase transition ($\Delta T_c=1.30$~K) and large residual resistivity ratio (RRR~=~38) indicate high quality of the film. Unlike single crystals with similar stoichiometry, $T_c$ of the thin film with optimal potassium concentration gradually increases to 40.8~K at 11.8~kbar ($P_c$) and decreases with further increasing pressure, showing a dome shape. The unusual pressure dependence of $T_c$ is ascribed to a lattice mismatch between the LAO substrate and thin film, where applied hydrostatic pressure effectively acts as uniaxial pressure along the crystalline c-axis of the film. The local temperature exponent of the resistivity reveals a funnel-shaped topology surrounding the optimal pressure $P_c$, indicating the presence of a quantum critical point where the normal-state properties are dominated by the associated quantum fluctuations.

This work was supported by Mid-career Researcher Program through NRF grant funded by the Ministry of Education, Science \& Technology (MEST) (No.~2010-0029136). TP acknowledges support from the NRF grant (No. 2010-0026762 \& 2011-0021645).


\begin{thebibliography}{30}

\bibitem{hosono08} Y. Kamihara, T. Watanabe, M. Hirano, and H. Hosono, J. Am. Chem. Soc. {\bf 130}, 3296 (2008).
\bibitem{stewart11} G. R. Stewart, Rev. Mod. Physics, {\bf83}, 1589 (2011).
\bibitem{clee08} C.-H. Lee, A. Iyo, H. Eisaki, H. Kito, M. T. Fernandez-Diaz, T. Ito, K. Kihou, H. Matsuhata, M. Braden, and K. Yamada, J. Phys. Soc. Jpn. {\bf77} 083704 (2008).
\bibitem{kuroki11} K. Kuroki, J. Phys. Chem. Solids {\bf72}, 307 (2011).
\bibitem{hiramatsu12} H. Hiramatsu, T. Katase, T. Kamiya, and H. Hosono, J. Phys. Soc. Jpn. {\bf81} 011011 (2012).
\bibitem{Iida09} K. Iida, J. Hnisch, S. Trommler, V. Matias, S. Haindl, F. Kurth, I. L. del Pozo, R. Hhne, M. Kidszun, J. Engelmann, L. Schultz, and B. Holzapfel, Appl. Phys. Lett. {\bf95}, 192501 (2009).
\bibitem{bellingeri10} E. Bellingeri, I. Pallecchi, R. Buzio, A. Gerbi, D. Marr, M. R. Cimberle, M. Tropeano, M. Putti, A. Palenzona, and C. Ferdeghini, Appl. Phys. Lett. {\bf96}, 102512 (2010).
\bibitem{slee09} S. Lee, J. Jiang, J. D. Weiss, C. M. Folkman, C. W. Bark, C. Tarantini, A. Xu, D. Abraimov, A. Polyanskii, C. T. Nelson, Y. Zhang, S. H. Baek, H. W. Jang, A. Yamamoto, F. Kametani, X. Q. Pan, E. E. Hellstrom, A. Gurevich, C. B. Eom, and D. C. Larbalestier, Appl. Phys. Lett. {\bf95}, 212505 (2009).
\bibitem{slee10} S. Lee, J. Jiang, Y. Zhang, C. W. Bark, J. D. Weiss, C. Tarantini, C. T. Nelson, H. W. Jang, C. M. Folkman, S. H. Baek, A. Polyanskii, D. Abraimov, A. Yamamoto, J. W. Park, X. Q. Pan, E. E. Hellstrom, D. C. Larbalestier, and C. B. Eom, Nat. Mater. {\bf9}, 397 (2010).
\bibitem{Iida10} K. Iida, S. Haindl, T. Thersleff, J. Hnisch, F. Kurth, M. Kidszun, R. Hhne, I. Mnch, L. Schultz, B. Holzapfel, and R. Heller, Appl. Phys. Lett. {\bf97}, 172507 (2010).
\bibitem{katase10} T. Katase, H. Hiramatsu, T. Kamiya, and H. Hosono, Appl. Phys. Express {\bf3}, 063101 (2010).
\bibitem{nlee10} N. H. Lee, S.-G. Jung, D. H. Kim, and W. N. Kang, Appl. Phys. Lett. {\bf96}, 202505 (2010).
\bibitem{takeda10} S. Takeda, S. Ueda, T. Yamagishi, S. Agatsuma, S. Takano, A. Mitsuda, and M. Naito, Appl. Phys. Express {\bf3}, 093101 (2010).
\bibitem{rotter08} M. Rotter, M. Pangerl, M. Tegel, and D. Johrendt, Angew. Chem. {\bf47}, 7949 (2008).
\bibitem{dmowski99} L. H. Dmowski and E. Litwin-Staszewska, Meas. Sci. Technol. {\bf10} 343 (1999).
\bibitem{yamazaki10} T. Yamazaki, N. Takeshita, K. Kondo, R. Kobayashi, Y. Yamada, H. Fukazawa, Y. Kohori, P. M. Shirage, K. Kihou, H. Kito, H. Eisaki, and A. Iyo, J. Phys.:Conf. Series {\bf215}, 012041 (2010).
\bibitem{torikachvili08} M. S. Torikachvili, S. L. Bud'ko, N. Ni, and P. C. Canfield, Phys. Rev. B {\bf78}, 104527 (2008).
\bibitem{luo08} H. Luo, Z. Wang, H. Yang, P. Cheng, X. Zhu, and H.-H. Wen, Supercond. Sci. Technol. {\bf21}, 125014 (2008).
\bibitem{zhao04} J. Zhao, N. L. Ross, and R. J. Angel, J. Phys.: Condens. Matter {\bf16}, 8763 (2004).
\bibitem{kimber09} S. A. J. Kimber, A. Kreyssig, Y.-Z. Zhang, H. O. Jeschke, R. Valent', F. Yokaichiya, E. Colombier, J. Yan, T. C. Hansen, T. Chatterji, R. J. McQueeney, P. C. Canfield, A. I. Goldman, and D. N. Argyriou, Nature Mat. {\bf8}, 471 (2009).
\bibitem{shen11} B. Shen, H. Yang, Z.-S. Wang, F. Han, B. Zeng, L. Shan, C. Ren, and H.-H. Wen, Phys. Rev. B {\bf84}, 184512 (2011).
\bibitem{kasahara10} S. Kasahara, T. Shibauchi, K. Hashimoto, K. Ikada, S. Tonegawa, R. Okazaki, H. Shishido, H. Ikeda, H. Takeya, K. Hirata, T. Terashima, and Y. Matsuda, Phys. Rev. B {\bf81}, 184519 (2010).
\bibitem{park08} T. Park, V. A. Sidorov, F. Ronning, J.-X. Zhu, Y. Tokiwa, H. Lee, E. D. Bauer, R. Movshovich, J. L. Sarrao, and J. D. Thompson, Nature {\bf456}, 366 (2008).

\end{thebibliography}
\end{document}